\documentstyle[12pt,epsf]{article}

\begin{document}

\newcommand{\qed}{\hfill$\Box$}
\newcommand{\ov}{\overline}
\newcommand{\proof}{{\bf Proof:}\ }
\newtheorem{theo}{Theorem}[section]
\newtheorem{conj}[theo]{Conjecture}
\newcommand{\head}[1]
{\markright{\hbox to 0pt{\vtop to 0pt{\hbox{}\vskip 3mm \hrule width
  \textwidth \vss} \hss}{\sc #1}}}

\title{Traveling Salesmen in the Presence of Competition} 
\author{S\'andor P. Fekete \\
Department of Mathematical Optimization\\
Braunschweig University of Technology\\
D--38106 Braunschweig, Germany\\
s.fekete@tu-bs.de\\
[3mm]
\and
Rudolf Fleischer\\
Department of Computer Science\\
Hong Kong University of Science and Technology\\
Clear Water Bay, Kowloon, Hong Kong\\
rudolf@cs.ust.hk\\
[3mm]
\and
Aviezri Fraenkel\\
Department of Computer Science and Applied Mathematics\\
Weizmann Institute of Science\\
76100 Rehovot, Israel\\
fraenkel@wisdom.weizmann.ac.il \\
[3mm]
\and
Matthias Schmitt\\
Center for Parallel Computing\\
Universit\"at zu K\"oln\\
D -- 50931 K\"oln, Germany\\
mschmitt@zpr.uni-koeln.de
}
\date{}

\pagestyle{myheadings}
\head{The Competing Salesmen Problem}

\thispagestyle{empty}
\maketitle

\vspace{3ex}
\begin{abstract}
We propose the
``Competing Salesmen Problem'' (CSP),
a 2-player competitive version of the
classical Traveling Salesman Problem.
This problem arises when considering
two competing salesmen instead of just one.
The concern for a shortest tour is replaced
by the necessity to reach any of the customers
before the opponent does. 

In particular, we consider the situation
where players take turns, moving along one
edge at a time within a graph $G=(V,E)$.
The set of customers is given by a subset
$V_C\subseteq V$ of the vertices. At any given time,
both players know of their opponent's position. A player
wins if he is able to reach a majority of the vertices
in $V_C$ before the opponent does. 

We prove that the CSP is PSPACE-complete, even if the graph is
bipartite, and both players start at distance 2 from each other.
Furthermore, we show that the starting
player may not be able to avoid losing the game, even if both players
start from the same vertex.
However, for the case of bipartite graphs, we show that the
starting player always can avoid a loss.
On the other hand, we show that the second
player can avoid to lose by more than one 
customer, when play takes place on a graph that is a tree $T$, and
$V_C$ consists of leaves of 
$T$. It is unclear whether a polynomial strategy
exists for any of the two players to force this
outcome.
For the case where $T$ is a star (i.\,e., a tree with only
one vertex of degree higher than two) and $V_C$
consists of $n$ leaves of $T$, we give a simple
and fast strategy which is optimal for both players.
If $V_C$ consists not only of leaves, we point out that
the situation is more involved.

\end{abstract}

\vspace{1cm}
{\bf Keywords:} Combinatorial games, complexity, PSPACE-completeness, strategy stealing,
Traveling Salesman Problem (TSP), Competing Salesmen Problem (CSP).

\vspace{1cm}
{\bf Classification:} 68Q25, 90D43, 90D46

\newpage
\section{Introduction}
\label{se:intro}
In the classical Traveling Salesman Problem (TSP),
we are given a (weighted) graph $G=(V,E)$ and the
task to find a shortest roundtrip that visits every vertex precisely once.
This reflects the situation where a salesman has to visit
a set of customers and return to his initial position.

However, 
a salesman may be confronted with competitors
who are eager to sign up the same clientele -- giving
a new twist to the old motto ``first come, first serve''.

This situation motivates the {\em ``Competitive Salesmen problem''}
(CSP): 

We are given a (directed or undirected) road system,
i.e., a graph $G=(V,E)$ 
and the locations of the customers, i.e.,
a subset $V_C\subseteq V$ of the vertices. 
There are two players, I and II, with starting positions
$v_I$ and $v_{II}$, and an initial score of zero. 
Starting with I, both players take turns
moving by changing from the current location to an 
adjacent vertex. Depending on the scenario, players may or may not
be allowed to pass. At any given time,
both players know of their own and their opponent's position
as well as about all the remaining vertices with customers.
If a player reaches a vertex with a customer,
his score is increased by one, and the vertex is removed from
the set $V_C$ of still available customers, but 
not removed from $V$. 
The game ends when no further customers can be captured,
i.\,e., when $V_C=\emptyset$ or when no player has a path to an uncaptured
customer. Whoever has a higher score at the end
of the play, wins. If both players end up with the same score,
the game is tied.

An immediate generalization is to consider two competing teams of
salesmen; in the CSP$(h,k)$, a move of player I consists of 
moving one of his $h$ pieces, while player II has the choice between
one of his $k$ pieces.

\newpage
\section{Preliminaries}
\label{se:prelim}
The CSP is a {\em combinatorial game}. 
See~\cite{Conway, Guy} for classical references on this well-studied area,
and~\cite{FG,FY,GR,Schaefer} for other related papers.
Here we just note an important distinction for the outcome
of games that are not won by either player:

A game that is won neither by I nor by II is called
\begin{itemize}
\item {\em tied}, if it ends with both players having the same score,
\item {\em drawn}, if it does not end.
\end{itemize}

Throughout this paper, we mostly concentrate on the case of an
undirected graph. Some of the results include the directed case,
but we do point out some additional difficulties in one interesting
case. Throughout the paper, there are a number of illustrations;
in these figures, the set $V_C\subseteq V$ is indicated by circled
dots. Without loss of generality, we assume that a start vertex 
never belongs to $V_C$;  in several cases, a start vertex
is indicated by a hexagon.

The rest of this paper is organized as follows. In Section~\ref{se:complexity},
we show that the CSP is PSPACE-complete, even for the case of bipartite
undirected graphs, with both players starting at distance 2 from each other.
In Section~\ref{se:common}, we discuss the situation in which both players start
from the same vertex. We show that even in this case, player I may not be
able to avoid a loss, and that there may be draws. We also show that in the case of bipartite graphs, player I can avoid a loss. We also show that this result 
does not apply to directed graphs. In Section~\ref{se:trees}, we give
some results and open problems for the special case of trees, and 
Section~\ref{se:stars} considers the further restriction to trees with only
one vertex of degree higher than two.

\newpage
\section{Complexity}
\label{se:complexity}
While the TSP on directed or undirected graphs is merely 
NP-complete, the two-player competitive game CSP turns out to be
PSPACE-complete.

\begin{theo}
\label{th:pspace}
The decision problem whether player~{\rm I} can win in {\rm CSP$(1,1)$} 
is PSPACE-complete, even for the special case of bipartite graphs,
with both players starting at distance 2 from each other. 
\end{theo}

\begin{proof}
A {\it position\/} in CSP$(h,k)$ is a quintuple 
$(\tau, G,V_C^{'},u_I,u_{II})$, 
where $\tau\in\{I,II\}$ indicates whether Player~I or Player~II moves 
from the position, $G=(V,E)$ is the (di-)graph on which the game is played, 
$V_C^{'}$ is the current set of uncaptured customers; and $u_I$ and $u_{II}$ 
are the vertices on which player~I and player~II reside. A draw can be 
declared after a position is repeated, that is, when a new position 
is encountered which is identical to a previous one. Identical positions 
can be detected by sequentially storing all the positions from the 
position at which $V_C^{'}$ was last diminished until it decreases again, 
beginning with the original $V_C$. If there are $h$ and $k$ salesmen for 
the two sides, at most $O(n^{h+k})$ positions have to 
be stored between any two consecutive changes of $V_C^{'}$, where $n=|V|$. 
If $h$ and $k$ are fixed, this is of polynomial size in the input size, 
in particular, if $h=k=1$. Therefore, for fixed $h$ and $k$, 
CSP$(h,k)\in $ PSPACE.

To see that CSP$(1,1)$ is PSPACE-hard, we describe 
a reduction from Quantified 3SAT (Q3SAT), where the Boolean 
formula $F$, containing $m$ clauses and $n$ variables, is in conjunctive 
normal form with 3 literals per clause. 

For technical reasons, and without loss of generality,
we shall make the following assumptions. 

\begin{itemize}
\item[(1)]~~The number $n$ of variables is even; because otherwise we may 
postfix $F$ with $\forall x_{n+1}\exists x_{n+2}\forall x_{n+3}\ 
(x_{n+1}+\ov{x}_{n+1}+x_{n+2})(\ov{x}_{n+2}+x_{n+3}+\ov{x}_{n+3})$. 
\item[(2)]~~There is a clause which contains a true literal and a false 
literal for every truth assignment of the variables; because if such a 
clause does not exist, then we can postfix $F$ with $\exists x_{n+1}\forall 
x_{n+2}\ (x_{n+1}+\ov{x}_{n+1}+x_{n+2})(x_{n+1}+x_{n+2}+\ov{x}_{n+2})$.
\end{itemize}\medskip

From a given instance of Q3SAT, we construct an instance of CSP$(1,1)$ by 
specifying the graph $G=(V,E)$ on which it is played. For simplicity
and clearer drawings, we first construct a graph that contains
some odd cycles in the subgraphs representing the clauses. 
It is straightforward to turn this graph into a bipartite one,
by subdividing all the original edges, in effect doubling all distances.

We proceed to describe $G$ by listing all vertices and edges of the
construction. In the following, we use $B:= n^2/2$ for simpler notation.
Different parts of the construction are shown in Figure~\ref{fi:variables}
for the variable gadget, in Figure~\ref{fi:clause} for the $m$ clause
gadgets, and in Figure~\ref{fi:cache} for a cache gadget.

\begin{eqnarray*}
V&=&\{x_i,\ov{x}_i : 1\le i\le n\}\\
&\cup&\{v_{I},v_{II}\}\\
&\cup&\{u_{i,h}: -1\le i\le n-2, 1\le h\le 2n\}\\
&\cup&\{u_{i,h}: n-1\le i\le n, 1\le h\le B\}\\
&\cup&\{v_j,a_j,b_j,c_j,y_j^1,y_j^2,y_j^3 : 1\le j\le m\}\\
&\cup&\{v_{0}\}\\
&\cup&\{p^k_{j,h}: 1\le k\le 3, 1\le j\le m, 1\le h\le B-n\}\\
&\cup&\{q_{i,h}: 0\le i\le 2n, 1\le h\le n^3\}\\
&\cup&\{d_i : 0\le i\le 5m+n-6\}. \\
& & \\
E&=&\{(v_I,v_{II}),(v_I,u_{-1,1}),(v_{II},u_{0,1})\}\\
&\cup&\{(u_{i,h},u_{i,h+1}) : -1\le i\le n-2, 1\le h\le 2n\}\\
&\cup&\{(u_{i,h},u_{i,h+1}) : n-1\le i\le n, 1\le h\le B\}\\
&\cup&\{(u_{i,2n},x_{i+2}),(u_{i,2n},\ov{x}_{i+2}) : -1\le i\le n-2\}\\
&\cup&\{(x_i,u_{i,1}),(\ov{x}_i,u_{i,1}) : 1\le i\le n\}\\
&\cup&\{(u_{2i+1,h},u_{2i+2,h}) : -1\le i\le n/2-1, 1\le h\le 2n\}\\
&\cup&\{(u_{n-1,h},u_{n,h}) : 1\le h\le B\}\\
&\cup&\{(u_{n-1,B},v_{0})\}\\
&\cup&\{(u_{n,B},a_{j}),(v_0,v_j),(a_j,b_j),(b_j,c_j): 1\le j\le m\}\\
&\cup&\{(y^1_j,y^2_j),(y^1_j,y^3_j),(y^2_j,y^3_j): 1\le j\le m\}\\
&\cup&\{(v_j,y^k_j),(c_j,y^k_j): 1\le k\le 3, 1\le j\le m\}\\
&\cup&\{(y^k_j,p^k_{j,1}): 1\le k\le 3, 1\le j\le m\}\\
&\cup&\{(p^k_{j,h},p^k_{j,h+1}): 1\le k\le 3, 1\le j\le m, 1\le h\le B-n\}\\
&\cup&\{(p^k_{j,B-n},x_{i}) : {\rm \ iff\ } c_j\ {\rm \ contains\ } x_i
{\rm \ as\ literal\ } k,\\
& & \ \ \ \ 1\le k\le 3,\ 1\le i\le n, 1\le j\le m\}\\
&\cup&\{(p^k_{j,B-n},\ov{x}_{i}) : {\rm \ iff\ } c_j\ {\rm \ contains\ } 
\ov{x}_i {\rm \ as\ literal\ } k,\\
& & \ \ \ \ 1\le k\le 3,\ 1\le i\le n, 1\le j\le m\}\\
&\cup&\{(x_i,q_{2i,1}),(\ov{x}_i,q_{2i-1,1}),(x_i,d_{0}),(\ov{x}_i,d_{0}) : 1\le i\le n\}\\
&\cup&\{(q_{i,h},q_{i,h+1}) : 0\le i\le 2n, 1\le h\le n^3-1\}\\
&\cup&\{(d_0,q_{0,1})\}\\
&\cup&\{(q_{i,n^3},d_1) : 0\le i\le 2n\}\\
&\cup&\{(d_{h},d_{h+1}) : 1\le h\le 5m+n-6\}.\\
\end{eqnarray*}
We single out the subset of $V$ on which customers reside: 
\begin{eqnarray*}
V_C&=&\{x_i,\ov{x}_i : 1\le i\le n\}\\
&\cup&\{a_j,b_j,y_j^1,y_j^2,y_j^3  : 1\le j\le m\}\\
&\cup&\{u_{i,h}: -1\le i\le n-2, 1\le h\le 2n\}\\
&\cup&\{u_{i,h}: n-1\le i\le n, 1\le h\le B\}\\
&\cup&\{d_i : 0\le i\le 5m+n-6\}. 
\end{eqnarray*}

\begin{figure}[htbp]
   \begin{center}
   \epsfxsize=.80\textwidth
   \ \epsfbox{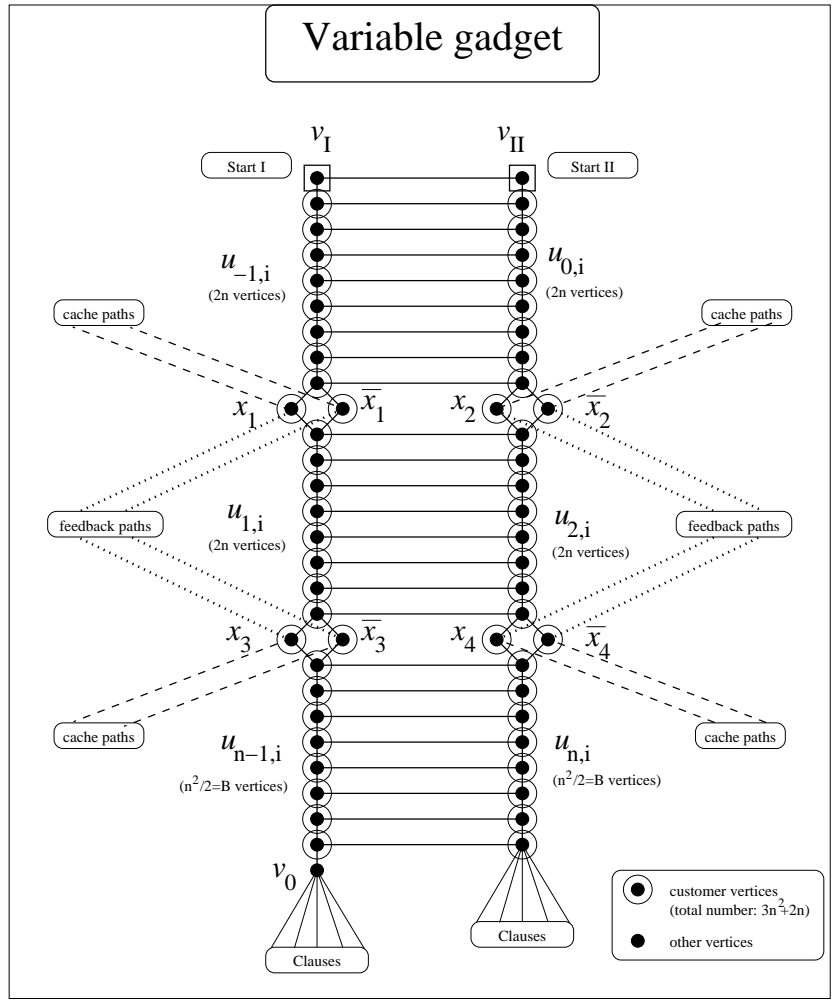}
   \caption{The variable gadget: Player I chooses a truth
    setting for the odd variables by running from $v_{I}$ to
    $u_{n-1,B}$, while player II chooses a truth setting for the even
    variables by running from $v_{II}$ to $u_{n,B}$.}
   \label{fi:variables}
   \end{center}
\end{figure}

\begin{figure}[htbp]
   \begin{center}
   \epsfxsize=.80\textwidth
   \ \epsfbox{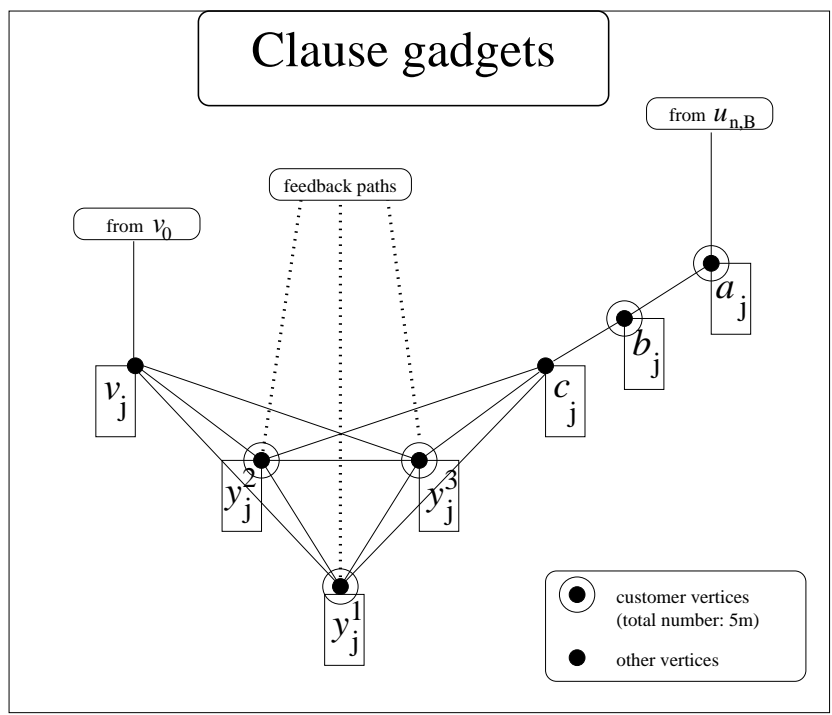}
   \caption{A clause gadget: Player II picks up two customers
    at $a_j$ and $b_j$, while player I collects two of the 
    customers at $y_j^k$, leaving the third $y_j^k$ to player II.
    The outcome of the game is decided by the possibility of picking
    up an extra customer on a variable, after traveling back to
    the variable gadget along a feedback path.}
   \label{fi:clause}
   \end{center}
\end{figure}

\begin{figure}[htbp]
   \begin{center}
   \epsfxsize=.80\textwidth
   \ \epsfbox{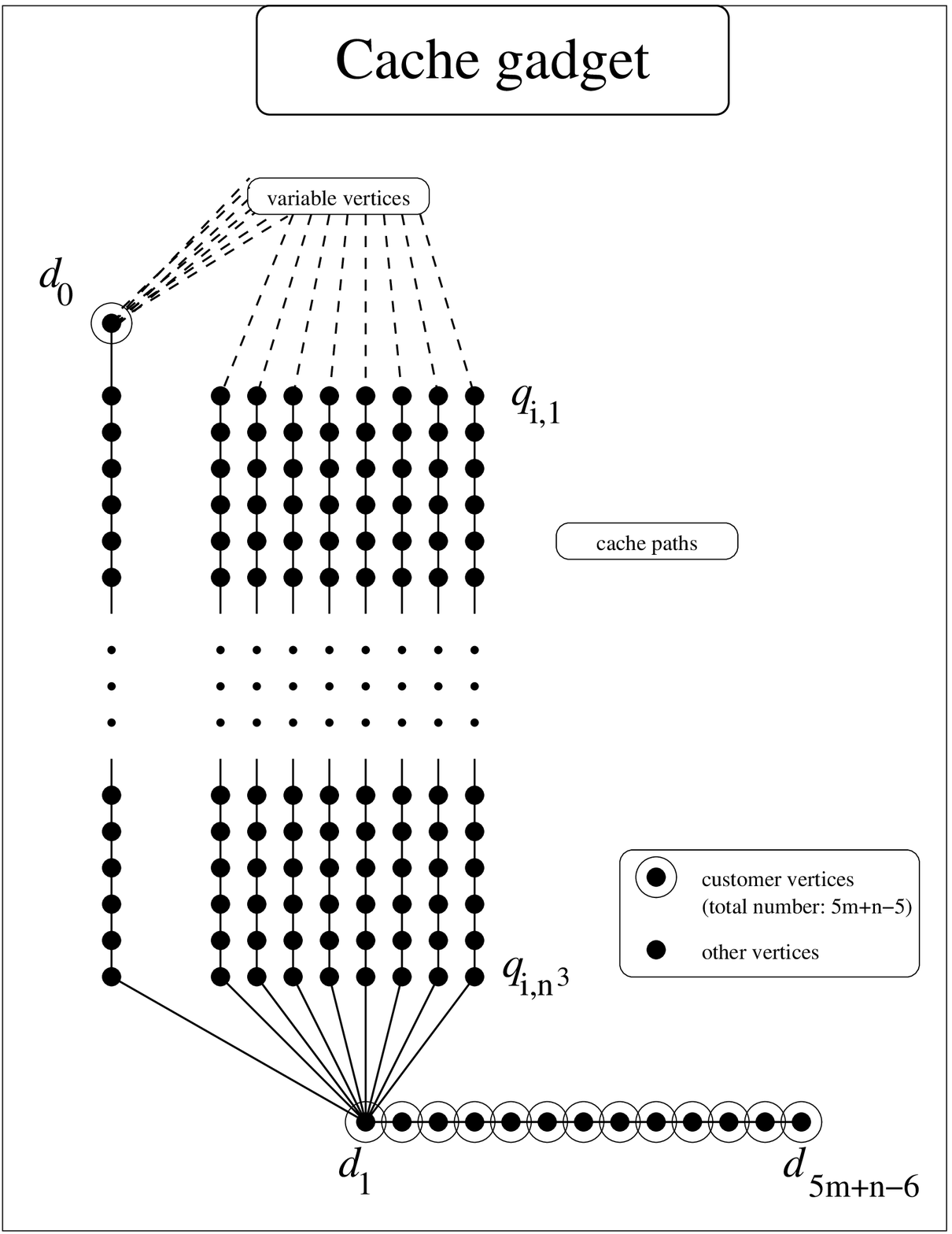}
   \caption{The cache gadget: A set of $5m+n-6$ customers very far from
    the rest of the graph, allowing a player to claim the victory if he
    has collected enough customers on the main part of the graph.
    The additional node $d_0$ breaks a tie in favor of player I iff
    player I wins the corresponding instance of Q3SAT.}
   \label{fi:cache}
   \end{center}
\end{figure}

\medskip
The initial vertices for player~I and player~II are $v_{I}$ and $v_{II}$, 
respectively. Note that $|V|=2n+2+2n^2+2B+7m+1+3m(B-n)+2n^4+5m+n-5=
2n^4+3mn^2/2+3n^2-3mn+12m+3n-4$, $\ |V_C|=2n+5m+2n^2+2B+5m+n-5=3n^2+10m+3n-5$. 
The construction is clearly polynomial. An example with $n=4$ is 
depicted in Figure~\ref{fi:variables}. (Note that this yields
$2n=8=n^2/2=B$.) A clause gadget is shown in Figure~\ref{fi:clause}.
To avoid cluttering the figure, some of the edges 
connecting the {\it diamonds\/} (the subgraphs induced by 
$(u_{i-2,2n},x_i,\ov{x}_i,u_{i,1})$) with other parts of the construction
are only shown symbolically. Figure~\ref{fi:cache} shows the structure of
the remaining part.
The vertices $p^k_{j,h}$ induce a collection of {\em feedback paths} 
that connect
the {\em triangles} $(y^1_j,y^2_j,y^3_j)$ in the gadget representing
clause $j$ to the variable representations of the literals
$y^1_j$, $y^2_j$, and $y^3_j$. 
\medskip

It will be useful to designate the subgraph of $G$ induced by 
$q_{0,1},\ldots,q_{2n,n^3}$, $d_0,d_1,\dots,d_{5m+n-6}$ as the 
{\it cache part\/} $G_2=(V_2,E_2)$ of 
$G$, and the subgraph of $G$ induced by $V\setminus V_2$ as the 
{\it main part} $G_1=(V_1,E_1)$.\medskip 

{\bf Remarks}.

\begin{itemize}
\item[(a)]~~We have $|V_C\cap V_2|=5m+n-6$, $|V_C\cap V_1|=3n^2+5m+2n+1$,
and a player wins by collecting at least $3n^2/2+5m+3n/2-2$ customers. 
\item[(b)]~~Moving from the main part to the cache part takes longer than
visiting all vertices of the main part.
\item[(c)]~~In the proof we shall see that in every play on $G$, all 
the customers are captured. Only after capturing at least $3n^2/2+n/2+4$ 
customers in the main part can a player win by moving to the cache
part. For by (a), $(3n^2/2+n/2+4)+5m+n-6=3n^2/2+5m+3n/2-2
=\lfloor |V_C/2|\rfloor +1$. 
Thus, a player who captured precisely $3n^2/2+n/2+4$ customers in the 
main part and then goes for the cache part before the opponent 
starts to do so wins by at least one customer. 
\end{itemize}

Here is the ``regular'' play on $G$. We shall see later 
that ``small'' deviations are compatible with regular play, but 
``large'' deviations lead to the defeat of the deviator. 

The players move down their respective sides of the ``ladders''
formed by the $u_{i,h}$, and traverse the diamonds according to 
their chosen truth assignments in the given instance of Q3SAT, 
i.e., traversing $x_i$ if $x_i=1$, otherwise traversing $\ov{x}_i$. 
In this manner, player I gets to assign the ``odd'', i.e.
existentially quantified, variables, while II gets to assign the 
``even'', i.e., all-quantified, variables.
After this stage, I makes the move 
$(u_{n-1,B},v_{0})$. Player~II now selects a clause $c_j$ by moving 
from $x_n$ to $a_j$. This is matched by the move $(v_{0},v_j)$ of 
I. While II traverses $b_j, c_j$, player~I selects the literal within $c_j$ 
to which II should move. Player~I enforces this by capturing the other
two of the customers in the triangle $T_j=(y_j^1,y_j^2,y_j^3)$. 
Then II captures the remaining $y_j^{k_3}$. Following the feedback path
incident to $y_j^{k_2}$, I takes $B-n$ moves to get to
a suitable literal, say $z_{i_2}$, on one of the diamonds, 
and II takes the same number of moves from $y_j^{k_3}$ along its feedback path
to a literal, say $z_{i_3}$, on a diamond. 

At this stage each of the players has captured $3n^2/2$
customers along the ladders, and $n/2$ customers on the 
diamonds; I has also captured two customers on the triangle $T_j$, and 
possibly a customer on $z_{i_2}$. Thus I has captured $3n^2/2+n/2$ 
plus 2 or plus 3 customers. 
Player~II has captured two customers on $a_j,b_j$ and one 
on $T_j$, possibly also one on $z_{i_3}$. Thus II has now captured $3n^2+n/2$ 
plus 3 or plus 4 customers. 

Suppose first that II can win in the given instance of Q3SAT. This means 
that for any truth assignment of the variables of player~I, player~II can 
assign truth values so that at least one clause $c_j$ is false, i.e, all 
the literals in $c_j$ have value 0. In terms of regular 
play, this means that II captured a customer on $z_{i_3}$. Thus II captured 
$3n^2/2+n/2$ plus 4 customers, at least one more than player~I. 
At this point I moves next. If I chooses to move straight to the cache
part, player I will capture a total of at most
$3n^2/2+n/2+3+5m+n-6=3n^2/2+5m+3n/2-3<|V_C|/2$. 
By restricting himself to the main 
part, II can capture all the rest, namely $3n^2/2+5m+3n/2-2$, 
so II wins by precisely one customer. 
Therefore player~I will make some other
move, e.g., the move $(z_{i_2},d_0)$. 
Then II responds by moving to the cache part. By 
Remark~(c), II thus wins in the constructed instance of CSP$(1,1)$ by 
precisely one customer. 

Secondly, suppose that player~I can win in the given instance of Q3SAT. 
This implies that player~I can assign truth values such that for any truth 
assignment of II, every clause contains at least one true literal. In terms 
of regular play on the constructed instance of CSP$(1,1)$, this means that 
player~I can arrange that $z_{i_3}$ will already have been captured during the 
initial diamond traversal, so II will have captured only $3n^2/2+n/2+3$ 
customers up to and including the feedback edge traversal. We consider 
two cases. 

\begin{itemize}
\item[(i)]~~The clause $c_j$ selected by II contains a false literal, 
say $z_{i_\ell}$. Then I continues according to regular 
play, playing in the triangle $T_j$ such that II is ``forced'' to 
move to $z_{i_k}$, and I himself moves to $z_{i_\ell}$. Then player~I will 
have captured $3n^2/2+n/2+3$ customers, the same as II. It is now the turn 
of I. As above it is seen that if I moves immediately to the cache
part, he loses by one customer. A better move for I is $(z_{i_\ell},d_{0})$. 
Now II cannot afford to let I take the cache,
so II has to move $(z_{i_k},q_{i_k,1})$. This allows I to take a remaining
customer on a literal vertex, maintaining a distance of $n^3$ to the cache.
II still has to guard the cache by limiting his distance from $d_1$ to at most
$n^3-1$, and is thus forced to move $(q_{i_k,1},q_{i_k,2})$.
Eventually, I picks up all remaining literal customers by going
via $d_0$. In a similar manner, I can collect all remaining customers
in the main part:
First I picks up all remaining customers at vertices $y_j^k$, which have
distance $n^3+B-n$ from the cache. This is possible without exceeding this
distance; at the same time, II cannot afford to move to the same distance
from the cache, as this would leave the cache unguarded.
Next, I can move on to collecting the vertices $b_j$ (which have distance
$n^3+B-n+2$ from the cache) one by one, and finally collect the vertices
$a_j$, which have distance $n^3+B-n+3$. At this point, I has won the game.

\item[(ii)]~~The clause $c_j$ selected by II contains no false literal. 
Then I deviates slightly from regular play, by moving to 
a clause, say $c_{\ell}$, which does contain a false literal $z_{i_f}$. 
Such a clause exists by assumption~(3) above. After I captures two customers 
in $T_{\ell}$, I moves to $z_{i_f}$, having thus far captured 
$(3n^2/2+n/2)+3$ customers. Then player~I continues as in the case (i),
winning. Note that 
the 3 paths leading out from $c_j$ to the diamonds all end in literals 
whose customers have already been captured, and that I wins independently 
of whether II captures one, two or three customers in $T_j$. 
\end{itemize}

We have shown that if the players stick to regular play, then player~I 
can win in Q3SAT if and only if player~I can win in CSP$(1,1)$. It 
remains to check nonregular play.\medskip 

First of all, not proceeding down a ladder (say, to collect a customer
at $d_0$ or at an additional literal vertex) takes at least two moves per 
additional customer. This lets the other player change over to the
deviator's side of the ladder, continue in a zig-zagging fashion back
and forth between both sides of and down the ladder, and thus continue
to collect one customer per move. Therefore, the violator loses in the balance,
compared to regular play.
Furthermore, remark (c) above implies that if either 
player goes for the cache part 
at any point during the diamond traversal before having traversed a 
feedback edge, then that player loses, since a minimum of $3n^2/2+n/2+4$ 
captures have to be made by a winning player in the main part prior 
collecting the cache. 

If II loses in CSP$(1,1)$, II can possibly use a feedback edge leading back 
to a literal $z_{i_\ell}$ whose customer has not been captured during 
the diamond traversal, by using a vertex of $T_j$ whose customer was 
already captured by I. But then the balance of customers captured by 
II up to the capture at $z_{i_\ell}$ is unchanged, and I still wins. 

Conversely, if II wins in CSP$(1,1)$, player~I might traverse a feedback 
edge after having captured only one customer on a triangle $T_{j'}$ 
(possibly $j'\ne j$), hoping to capture enough customers 
during a second traversal of the diamonds, before II will have captured 
enough. An easy accounting argument, left to the reader, shows that 
player~I cannot muster a sufficient supply of customers with this maneuver. 
\end{proof}\qed\bigskip

It is not hard to see that the above construction can be modified
to establish a proof of the PSPACE-completeness of the CSP
on bipartite directed graphs without antiparallel edges.
Furthermore, it can be modified to cover the scenario in which
both players move simultaeneously: after scaling edge lengths by a factor 
of two, give Player I a headstart of one move.

\section{Identical Starting Point}
\label{se:common}
The result in the previous section shows that deciding the outcome
of a CSP instance is quite difficult, even when both players start very close
to each other, and the graph is bipartite. In this section, we concentrate
on the natural special case in which both players start from the 
{\em same} vertex. As it turns out, this scenario is quite different.

\begin{theo}
\label{th:bipartite}
For the CSP on bipartite graphs with both players starting at the same point, 
player I can avoid a loss.
\end{theo}

\begin{proof}
By way of contradiction, suppose player II has a winning strategy, 
and both players
start from vertex $v_0$. Now suppose I moves to vertex $v^I_1$, and   
II can counter this move by $v^{II}_1$. In the following,
let I visit any sequence of vertices $v_1^I,v_2^I,v_3^I,\ldots$
By assumption, II has a winning strategy, so there is a sequence of
moves $v_1^{II},v^{II}_2,v^{II}_3,\ldots$ that ends with II capturing an
absolute majority of customers. Note that each $v_i^{II}$
is determined by the sequence $v_1^I,v_2^I,v_3^I,\ldots,v_i^I$;
by induction, we can write $v_i^{II}(v_i^I)$ to indicate that I's move
to $v_i^I$ was successfully countered by II by moving to
$v_i^{II}$.

Now consider, for any sequence $u^{II}_1,u^{II}_2,u^{II}_3,\ldots$
of moves by II, the following sequence of moves for I:
\[v_1^I,v_0,v_1^{II}(u_1^{II}),v_2^{II}(u_2^{II}),v_3^{II}(u_3^{II}),\ldots\]
This means that player I gives up two moves by
moving to any neighbor $v_1^I$ and back to $v_0$, thereby giving player II
a head start of 1 1/2 moves. Then I plays against a ``phantom player'' 
II' that is one move lagging behind the real player II, 
i.e., precisely 1/2 move ahead of I, which allows I to ``steal''
the assumed second player's strategy against such a player.

By assumption, the above is a well-defined sequence of moves for I. 
Therefore, player I wins more than half of the customers against II', but
II wins more than half of the customers against I. Hence, there must
be a customer (say, at vertex $v_*$) that I reaches before II',
but that II reaches before I. Therefore, I must
be moving to vertex $v_*$ when II is already there, and just before
II' gets there.

This is a contradiction to the bipartiteness of $G$: After a move of
I, both players must always occupy vertices of opposite color, so
$I$ cannot reach $v_*$ with II positioned on that vertex.

Therefore, II cannot have a winning strategy, proving the claim.
\end{proof}\qed\bigskip

Note that the theorem remains valid for directed graphs,
if there is a single pair of anti-parallel edges that allows
I to leave and return to $v_0$ in just two moves.

The following example (courtesy of David Wood) shows that the possibility
of moving back to $v_0$ is crucial for the proof: In the absence of
an undirected edge at $v_0$, player I may be limited to winning a single
customer.

\begin{theo}
\label{th:wheel}
There is a family of instances of CSP on directed
graphs in which I cannot win more than 1 out of $n$ customers.
\end{theo}

\begin{figure}[htbp]
   \begin{center}
   \epsfxsize=.44\textwidth
   \ \epsfbox{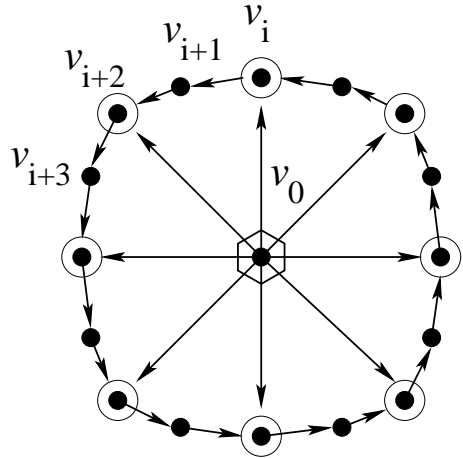}
   \caption{Player I loses by $n-1$ customers.}
   \label{fi:wheel}
   \end{center}
\end{figure}

\begin{proof}
Consider the graph shown in Figure~\ref{fi:wheel}. The initial
vertex for both players is denoted by $v_0$, the vertex set $V_C$ is
indicated by the circled vertices. 

Suppose I starts by moving to vertex $v_i$; then II responds by
moving to vertex $v_{i+2}$. Now the rest of the game is determined,
and I only wins the customer at $v_i$.
\end{proof}\qed\medskip

For non-bipartite graphs, I may lose, even on undirected graphs,
provided passing is not allowed.

\begin{theo}
\label{th:q.wins}
There are instances where I cannot avoid a loss,
even if both players start from the same vertex.
\end{theo}

\begin{proof}
Consider the graph shown in Figure~\ref{fig:q.wins}.
The initial vertex for both players is denoted by $v_0$,
the vertex set $V_C$ is indicated by the three circled vertices.

\begin{figure}[htbp]
   \begin{center}
   \epsfxsize=.44\textwidth
   \ \epsfbox{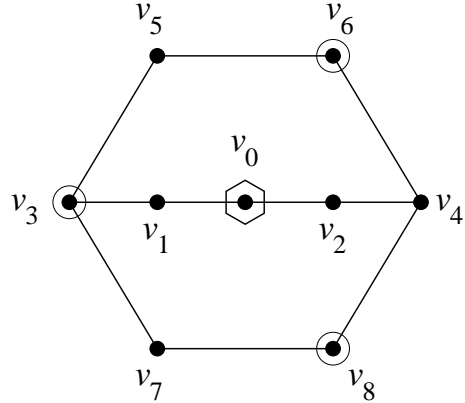}
   \caption{Player I loses.}
   \label{fig:q.wins}
   \end{center}
\end{figure}

Player I is in a zugzwang situation:
Suppose I starts by moving to vertex $v_1$;
then II moves to vertex $v_2$. If then I moves back
to $v_0$, II moves on to $v_4$; it is straightforward
to check that now II will force a win by taking the customer
at $v_6$ and at least one of $v_3$ and $v_8$.
Hence we can assume that I's second move is to $v_3$.
However, this is answered by II by moving to $v_4$,
and I cannot prevent II from taking both $v_6$ and $v_8$.

Therefore consider the case where I starts by moving
to $v_2$; II responds by moving to $v_1$. As in the
previous case, II wins by moving to $v_3$ if I moves
back to $v_0$. 
Hence we can assume that I's second move is to $v_4$,
followed by II moving to $v_3$, securing the first
customer. Regardless of I's next
move, II can again force a win by taking at least one
of the remaining two customers.

This concludes the proof.
\end{proof}\qed\medskip

An immediate consequence is the following:

\begin{theo}
\label{th:draw}
There are instances where optimal play from both I and II
forces a draw,
even if both players start from the same vertex.
\end{theo}

\begin{proof}
Consider the graph shown in Figure~\ref{fig:draw}.
The initial vertex for both players is denoted by $v_0$,
the vertex set $V_C$ is indicated by the three circled vertices.

\begin{figure}[htbp]
   \begin{center}
   \epsfxsize=.44\textwidth
   \ \epsfbox{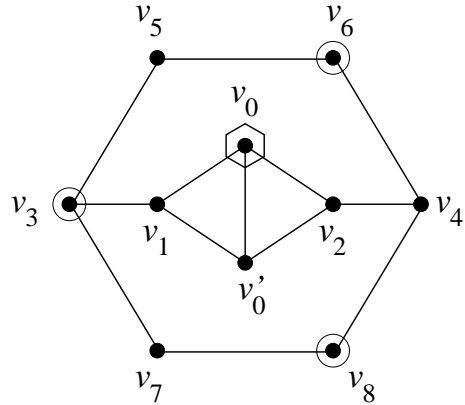}
   \caption{A draw game}
   \label{fig:draw}
   \end{center}
\end{figure}

Suppose player $X$ is the first to move to one of the vertices
in $\overline{V}=\{v_1,v_2\}$, while the other player $Y$ has not left
the vertex set $\{v_0, v_{0}'\}$. Then the analysis of Theorem~\ref{th:q.wins}
shows that player $Y$ can force a win by moving
to the other vertex in $\overline{V}$.

Therefore, neither of the players is willing to leave the set
$\{v_0, v_{0}'\}$, resulting in a draw.
\end{proof}\qed\medskip

\section{Trees}
\label{se:trees}
Our proof of Theorem~\ref{th:bipartite} is purely existential.
Furthermore, there is still no proof that player I cannot just avoid a loss,
but also avoid a draw, and end the game with a win or a tie.

\begin{conj}
\label{co:force}
For the CSP on trees, one of the players can force a win or a tie.
\end{conj}

In particular, this implies that for the case of identical starting 
point and an odd number of customers, player I always wins the game.

We have a pretty good idea how to tackle this problem; a constructive
argument may use bookkeeping on subsets of customers, and
the number of moves necessary to collect them and return to a previous position.
(This means generalizing the idea of playing against a phantom player,
and arguing that on trees it can only be an advantage to have extra
moves to spare.)
We hope to finish this argument at a later time. But even if this works out,
the resulting construction is exponential in size and rather awkward.
It would be a lot more satisfying to have a simple strategy that guarantees
a win for player I.

However, there are a number of difficulties that are indicated by the following
observations.

\begin{theo}
\label{th:vorsprung}
There are instances of CSP on trees, with both players starting
from the same vertex $v_0$, the number of customers being $2k+1$, and
the only way for player I to win allows II to potentially collect
$k$ customers before I reaches even a single customer.
\end{theo}

\begin{proof}
Consider the graph shown in Figure~\ref{fi:vorsprung}.
It has $k$ customers at an intermediate distance from the starting
vertex $v_0$, and at twice that distance from each other.
Furthermore, there is a cluster of $k+1$ customers at a large distance
from $v_0$, but at a small distance from each other.
 
\begin{figure}[htbp]
   \begin{center}
   \epsfxsize=.21\textwidth
   \ \epsfbox{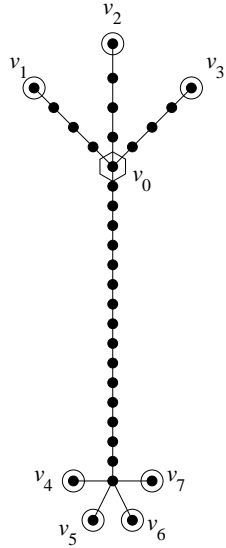}
   \caption{Player I can only win when accepting to trail by a large number.}
   \label{fi:vorsprung}
   \end{center}
\end{figure}

If I starts by taking one of the nearer customers, II collects
all the distant customers and wins. On the other hand, II may take
all the near customers before I takes one of the distant ones.
\end{proof}\qed\medskip

\begin{theo}
\label{apriori}
Consider instances of CSP on trees, where both players start
on the same vertex $v_0$. In general, player I cannot avoid
a loss by adapting an a-priori strategy, i.\,e., by prioritizing
the customers in an appropriate way, and always trying to collect
the customer with the highest priority.
\end{theo}

\begin{proof}
Consider the graph shown in Figure~\ref{fi:apriori}.
It consists of a symmetric tree with nine customers at the leaves,
grouped into three triples.

\begin{figure}[htbp]
   \begin{center}
   \epsfxsize=.75\textwidth
   \ \epsfbox{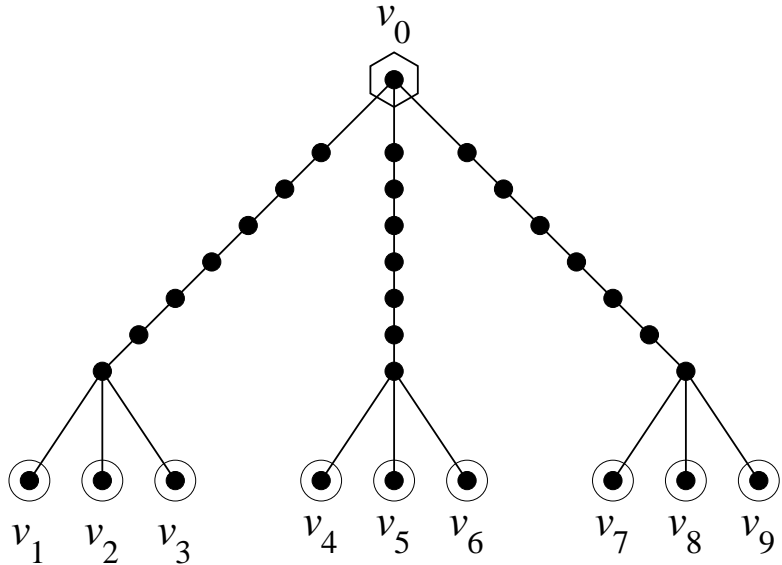}
   \caption{Player I loses when adapting an a-priori strategy.}
   \label{fi:apriori}
   \end{center}
\end{figure}

Without loss of generality, assume that $v_1$ is the first customer
on I's list; furthermore, assume that $v_4$ is the first customer
on the list that is none of $v_1$,  $v_2$, $v_3$.
Then it is straightforward to check that II can collect the customers
$v_5$, $v_6$, $v_7$, $v_8$, $v_9$ without any interference from I,
thus winning the game: In order to win, I would be forced to visit
customers from all three different clusters, which takes longer
than II needs to collect all five customers.
\end{proof}\qed\medskip

Finally, player II can limit his losses in a natural special case.
A rather involved argument for the following can be found in the 
fourth author's thesis. We omit this proof, as we believe that
there should be a relatively simple argument;
in particular, a proof of Conjecture~\ref{co:force} as described
should do the trick.

\begin{theo}
\label{q.tree}
Consider an instance of the CSP where the graph $G$ is a tree $T$
and both players start at the same vertex $v_0$. Suppose
all customers are positioned at leaves of the tree. Then 
player II can avoid a loss by more than one customer.
In particular, II can avoid a loss when the number of customers
is even.
\end{theo}

\section{Stars}
\label{se:stars}
It may be easier to come up with a strategy for the class of CSP instances
in which the graph $G$ is a star, i.e., a tree $T$ with at most one
vertex of degree higher than two. Again, we consider both players starting
from the same vertex. If customers are only contained in leaves, it is not hard to see 
that there is a simple optimal strategy for both players,
by always choosing
the nearest free customer when returning to the central node.

\begin{figure}[htbp]
   \begin{center}
   \epsfxsize=.44\textwidth
   \ \epsfbox{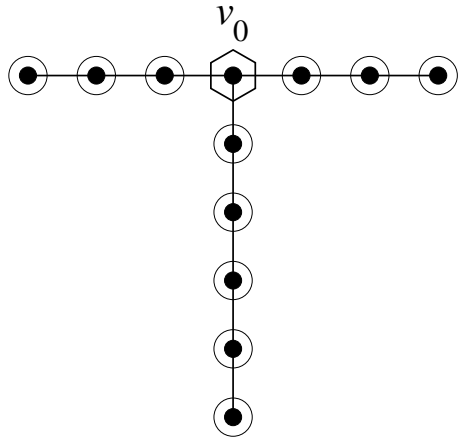}
   \caption{Can you give a {\em simple} winning strategy for player I,
and a short proof that it wins?}
   \label{fi:star}
   \end{center}
\end{figure}

However, it is straightforward to see that the example in Figure~\ref{fi:star}
cannot be won using this approach: If I collects all five customers
along the longest ray, II gets all customers along the other two rays.
Similarly, it follows that I loses when trying to pick up all the 
customers along a single ray. We leave it to the reader as an exercise
to work out a winning strategy for this instance.

\newpage
\section{Conclusions}
\label{se:conc}
In this paper we have introduced the Competing Salesman Problem.
Many open problems remain. Besides the ones mentioned directly or
indirectly throughout the paper, there are many more. One of the
more interesting scenarios may be a continuous geometric version,
in which customers are points in some space of fixed dimension, 
and the players move continuously or in discrete portions along 
arbitrary paths. Clearly, this introduces additional difficulties.

As there are innumerable variants of the TSP, due to many different
practical constraints, requirements, or objective functions, it is
quite conceivable that there are many more related games. One such variant
(called the freeze tag problem, FTP) has been considered in~\cite{FTP},
where a set of cooperating players have to awake each other, and any
awake player can awake a sleeping player by moving next to him.
In the original game of freeze tag, there are two competing teams,
and one wins if it can freeze all opposing players, while the second one
tries to avoid this permanently.

\section*{Acknowledgments} 
Some of this work was stimulated and motivated by the Dagstuhl Seminar
on Algorithmic Combinatorial Game Theory, 17--22 February, 2002,
organized by Erik Demaine, Rudolf Fleischer, Aviezri Fraenkel, and
Richard Nowakowski. We thank all participants for contributing to the fruitful
atmosphere of this workshop, in particular David Wood for pointing
out the example in Theorem~\ref{th:wheel}. 

Earlier parts of this
work originated from work on the fourth author's diploma 
thesis~\cite{Matze};
other parts of this paper were written during a stay at the Department
of Computer Science and Software Engineering, University of Newcastle,
NSW, Australia, where S\'andor Fekete was supported by a Visiting
Researcher Grant.
Rudolf Fleischer's work described in this paper
was partially supported by a grant
from the Research Grants Council of the Hong Kong Special Administrative
Region, China (Project No.~HKUST6010/01E).


\begin{thebibliography}{99}

\bibitem{FTP}
E.\ M.\ Arkin, M.\ Bender, S.\ P.\ Fekete,
J.\ S.\ B.\ Mitchell, M.\ Skutella.
The freeze-tag problem: How to wake up a swarm of robots.
{{\em Proc. 13th ACM-SIAM Symposium on Discrete Algorithms}}, 2002,
pp.\ 568--577.

\bibitem{Conway}
{E.\ R.\ Berlekamp, J.\ H.\ Conway, and R.\ K.\ Guy}.
{\em Winning ways for your mathematical plays}.
Academic Press, London, 1982.

\bibitem{FG}
{A.\ S.\ Fraenkel and E.\ Goldschmidt}.
PSPACE-Hardness of some combinatorial games.
{\em J.\ Combin.\ Theory (Ser.\ A)}, {\bf 46}, 1987, pp.\ 21--38.

\bibitem{FY}
A.\ S.\ Fraenkel and Y.\ Yesha.
Complexity of problems in games, graphs and algebraic equations. 
{\em Discrete Appl. Math.\/}, {\bf 1} (1979), pp.\ 15--30.

\bibitem{GR}
A.\ S.\ Goldstein and E.\ M.\ Reingold.
The complexity of pursuit on a graph. 
{\em Theoret. Comput. Sci. \emph{(Math Games)}\/}, {\bf 143} (1995), 
pp.\ 93--112.
 
\bibitem{Guy}
R.\ K.\ Guy (ed.).
{\em Proceedings of Symposia in Applied Mathematics}.
Vol.\ 43, Combinatorial Games.
American Mathematical Society, Providence, 1991.

\bibitem{Schaefer}
T.\ J.\ Schaefer.
On the Complexity of some two-person perfect-in\-for\-mation games.
{\em J.\ Comput.\ System Sci.}, {\bf 16}, 1978, pp.\ 185--225.

\bibitem{Matze}
{M.\ Schmitt}.
{\em Das Problem der konkurrierenden Handlungsreisenden 
    -- ein kombinatorisches Spiel}.
Diplomarbeit (Master's thesis).
Mathematisches Institut, Universit\"at zu K\"oln, 1997.

\end{thebibliography}
\end{document}